\documentclass[aps, prd, groupedaddress, preprint, tightenlines,  eqsecnum, nofootinbib, showpacs]{revtex4}
\usepackage{epsfig}
\usepackage{setspace}
\usepackage{graphicx}
\usepackage{leqno}
\usepackage{cancel}
\usepackage{amsmath}
\usepackage{amsfonts}
\usepackage{amssymb}
\usepackage{enumerate}

\usepackage{listings}

\usepackage{axodraw, float, slashed, graphicx, amssymb, amsmath}
\usepackage[usenames, dvipsnames]{xcolor}
\usepackage{booktabs}

\usepackage[margin=2.2cm, a4paper, includefoot]{geometry}

\def\equn{\refstepcounter{equation}\eqno({\rm \theequation})}

\def\spa#1.#2{\left\langle#1\,#2\right\rangle}
\def\spb#1.#2{\left[#1\,#2\right]}
\def\la{\langle}
\def\ra{\rangle}
\def\Mloop{M^{\oneloop}}
\def\Mtree{M^{\tree}}

\def\Aloop{A^{\oneloop}}

\def\coli#1#2{\mathop{\longrightarrow}^{#1 \parallel #2}}

\newcommand{\oneloop}{\text{1-loop}}

\newcommand{\tree}{\text{tree}}

\DeclareMathOperator{\SP}{\mathrm{Sp}}

\newcommand\figref[1]{fig.~\ref{#1}}
\def\eps{\epsilon}

\def\be{\begin{equation}}
\def\ee{\end{equation}}

\def\B{{B}}
\def\C{{C}}

\def\ceta{\eta}
\begin{document}

\hfill\today 

\title{$n$-point amplitudes with a single negative-helicity graviton}

\author{Sam~D.~Alston, David~C.~Dunbar and Warren~B.~Perkins}

\affiliation{
College of Science, \\
Swansea University, \\
Swansea, SA2 8PP, UK\\
\today
}

\begin{abstract}

We construct
an expression for the $n$-point one-loop graviton scattering
amplitude with a single negative helicity external leg using an augmented recursion  technique.
We analyse the soft-limits of these  amplitudes
and demonstrate that they
have soft behaviour beyond the conjectured universal behaviour.

\end{abstract}


\maketitle

\section{Introduction}

Exploring the singularities of scattering amplitudes has been a key
theme of research in recent years~\cite{RecentConference}  with the aim being to compute
amplitudes entirely from  a knowledge of their singular behaviour.  This reprise of earlier work~\cite{Eden} 
has been
invigorated both by developments in the understanding of underlying symmetries~\cite{Witten:2003nn} and technical progress. 

One important technique has been BCFW~\cite{Britto:2005fq} recursion which applies
complex analysis to amplitudes.  Using Cauchy's theorem, if a complex function is
analytic except at simple poles $z_i$ (all non-zero) and $f(z)\longrightarrow
0$ as $|z|\longrightarrow\infty$ then
by considering the integral
\begin{equation}
 \oint_C  f(z) {dz\over z}
\end{equation}
where the contour $C$ is the circle at infinity, we obtain  
\begin{equation}
f(0) =-\sum_i  { {\rm Residue}( f,z_i) \over z_i }
\end{equation}
The recursion procedure for amplitudes utilises this by complexifying two of the external 
momenta\footnote{As usual,  a null momentum is represented as a
pair of two component spinors $p^\mu =\sigma^\mu_{\alpha\dot\alpha}
\lambda^{\alpha}\bar\lambda^{\dot\alpha}$. For real momenta
$\lambda=\pm\bar\lambda^*$ but for complex momenta $\lambda$ and
$\bar\lambda$ are independent~\cite{Witten:2003nn}.  
We are using a spinor helicity formalism with the usual
spinor products  $\spa{a}.{b}=\epsilon_{\alpha\beta}
\lambda_a^\alpha \lambda_b^{\beta}$  and 
 $\spb{a}.{b}=-\epsilon_{\dot\alpha\dot\beta} \bar\lambda_a^{\dot\alpha} \bar\lambda_b^{\dot\beta}$.}
by shifting a pair of spinors
\begin{equation}
\bar\lambda_{{a}}\to \bar\lambda_{\hat{a}} =\bar\lambda_a - z \bar\lambda_b 
\qquad , \qquad
\lambda_{{b}}\to\lambda_{\hat{b}} =\lambda_b + z \lambda_a 
. 
\label{BCFWshift}
\end{equation}
This shifts the momenta $p_a$ and $p_b$ to complex values $p_a(z)$ and $p_b(z)$  which
are both still null and preserves overall momentum conservation.  The resultant amplitude $A(z)$ plays the role of $f(z)$ above.

As tree amplitudes are rational functions of
$\lambda_i$ and $\bar\lambda_i$, provided the complexified amplitude
$A(z)$ satisfies the conditions above, $A(0)$ may be
determined in terms of residues. 
These residues arise from simple poles corresponding to 
factorisations of the amplitude~\cite{Bern:1995ix}. The target amplitude is thus readily expressed in terms of lower point amplitudes
evaluated with specific complex momenta,
\begin{equation}
A_n^\tree (0) \; = \; \sum_{i,\lambda} {A^{\tree,\lambda}_{r_i+1}(z_i)
  {i\over K^2}A^{\tree,-\lambda}_{n-r_i+1}(z_i)},
\label{RecursionTree}
\end{equation}
where the summation over $i$ is only over factorisations where the $a$
and $b$ legs are on opposite sides of the pole. This is the on-shell
recursive expression of~\cite{Britto:2005fq}. 

One-loop amplitudes in a massless theory can be expressed as~\cite{Bern:1994cg}
$$
\Aloop_n=\sum_{i\in \cal C}\, a_i\, I_4^{i} +\sum_{j\in \cal D}\,
b_{j}\, I_3^{j} +\sum_{k\in \cal E}\, c_{k} \, I_2^{k} +R_n +O(\eps),
\equn\label{DecompBasis}
$$ 
where the $I_r^i$ are $r$-point scalar integral functions and the $a_i$
etc. are rational coefficients. $R_n$ is a purely rational term.
In terms of complex momentum this means the amplitude  has both poles and discontinuities.  Eq.\eqref{DecompBasis} may simplify
in specific theories, e.g. in maximal supersymmetric Yang-Mills and supergravity where only the 
first ''box functions'' appear~\cite{Bern:1994zx,BjerrumBohr:2006yw} and in specific helicity amplitudes such are the ``all-plus'' and ``single-minus'' amplitudes of both Yang-Mills and gravity where 
only the rational terms $R_n$ appear. 

The appearance of discontinuities is in fact enormously helpful in computing amplitudes.  These
discontinuities allow the direct computation of the coefficients of the scalar integral functions using unitarity methods. 
Dividing the amplitude into integral functions with rational
coefficients has been very fruitful: a range of specialised techniques
have been devised to determine the rational coefficients based
on unitarity techniques rather than Feynman diagrams~\cite{Cutkosky:1960sp,
Bern:1994zx,Bern:1994cg,%
Bern:1997sc, BrittoUnitarity}.
Progress has been made both via the two-particle cuts~\cite{Bern:1994zx,Bern:1994cg,Dunbar:2009ax}
and using 
generalisations of unitarity~\cite{Bern:1997sc} where, for example, 
triple~\cite{Bidder:2005ri,Darren,BjerrumBohr:2007vu,Mastrolia:2006ki} and quadruple cuts~\cite{BrittoUnitarity} 
are utilised to identify the triangle and box coefficients respectively.

Returning to the amplitude, there is a further problem related to complex factorisation.  In general, beyond tree level, amplitudes with complex momenta 
may have multi-pole singularities  and consequently may have poles of higher order. 
Mathematically, this is not a barrier to using complex analysis, since, if we have a function whose expansion about $z_i$ is 
$$
f(z) = {a_{-2} \over (z-z_i)^2 }+
{a_{-1} \over (z-z_i) } +{\rm finite} 
\equn
$$
then 
$$
{\rm Residue}( { f(z) \over z }, z_i)  =  -{a_{-2} \over z_i^2 } +{a_{-1} \over z_i} 
\equn
$$
However for one-loop amplitudes only the leading singularities have been determined in general and there are no
general theorems for the sub-leading terms. 

In one-loop amplitudes poles in momenta arise in two ways: firstly
from explicit poles in Feynman diagrams such as in
fig.~\ref{fig:AA},  and secondly from loop momentum
integrals. Specifically a $(P^2)^{-1}$ pole can arise from the loop
momentum integral of diagrams of the form shown in fig.~\ref{fig:BB}.  
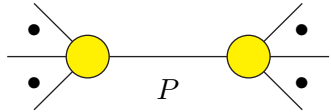
\begin{figure}[H]
\centerline{
    \begin{picture}(100,100)(0,-20)    
     \Line(60,20)(0,20)
     \Line(0,20)(-20,0)
     \Line(0,20)(-30,20)
     \Line(0,20)(-20,40)
     \Line(60,20)( 80,0)
     \Line(60,20)( 90,20)
     \Line(60,20)( 80,40)
     \CCirc(0,20){8}{Black}{Yellow} 
     \CCirc(60,20){8}{Black}{Yellow} 
     \Text(-20,30)[c]{$\bullet$}
     \Text(-20,10)[c]{$\bullet$}
     \Text( 80,30)[c]{$\bullet$}
     \Text( 80,10)[c]{$\bullet$}
     \Text(30,8)[c]{$P$}
    \end{picture}
    }
    \caption{Diagrams with an explicit pole.}  
    \label{fig:AA}
\end{figure}
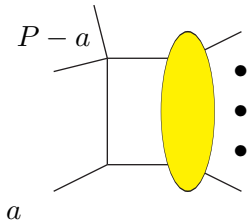
\begin{figure}[H]
\centerline{
    \begin{picture}(100,100)(0,-20)    
     \Line(20,0)(50,0)
     \Line(20,40)(50,40)
     \Line(20,0)(20,40)
    \Line(20,0)(0,-10)
    \Line(20,40)(0,35)
   \Line(20,40)(15,60)
\Line(50,40)(70,50)
\Line(50,0)(70,-10)
   \COval(50,20)(30,10)(0){Black}{Yellow} 
   \Text(-15,-18)[c]{$a$}
   \Text(70,35)[c]{$\bullet$}
   \Text(70,20)[c]{$\bullet$}
   \Text(70,05)[c]{$\bullet$}
   \Text(0,48)[c]{$P-a$}
    \end{picture}
    }
    \caption{Diagrams with a pole from loop integration.}  
   \label{fig:BB}\end{figure}

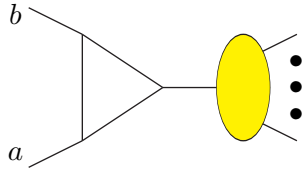
\begin{figure}[H]
\centerline{
    \begin{picture}(100,100)(0,-20)    
     \Line(20,0)(50,20)
     \Line(20,40)(50,20)
     \Line(20,0)(20,40)
    \Line(20,0)(0,-10)
    \Line(20,40)(0,50)
\Line(50,20)(90,20)
\Line(80,30)(100,40)
\Line(80,10)(100,0)
\COval(80,20)(20,10)(0){Black}{Yellow} 
  \Text(-5,48)[c]{$b$}
  \Text(-5,-5)[c]{$a$}
   \Text(100,30)[c]{$\bullet$}
   \Text(100,20)[c]{$\bullet$}
   \Text(100,10)[c]{$\bullet$}
    \end{picture}
    }
    \caption{Double poles may arise from these diagrams.}  
   \label{fig:CC}
   \end{figure}

When the pole is a two-particle factorisation $P^2=(k_a+k_b)^2$, these two sources can overlap as 
in  fig.~\ref{fig:CC} and produce double poles.

Figure~\ref{fig:CC}  illustrates the challenge in determining the sub-leading contribution.
Although the poles are physical and gauge independent, the diagram  effectively contains an off-shell current whose sub-leading
term in $s_{ab}$ is gauge and scheme dependent.   The leading term can be shown to be expressible in terms of an "effective vertex" 
times an on-shell tree amplitude~\cite{Bern:2005hs,Brandhuber:2007up}. The vertex vanishes unless both legs $a$ and $b$ have the same helicity in which case the double pole
is
$$
 V(a^+,b^+, P^+) \times { 1 \over s_{ab} } \times A^{\tree}(P^-, \cdots )
\equn
$$
The tree amplitude in this expression is the lower point amplitude where legs $a$ and $b$ have been replaced by 
a single leg of negative helicity.  In Yang-Mills theory the effective vertex takes the form
$$
 V^{YM}(a^+,b^+, P^+)  =  -{i  \over 48 \pi^2 }
 { \spb{a}.b \spb{b}.P \spb{P}.a   \over s_{ab} } \propto { \spb{a}.b \over \spa{a}.b }
\equn
$$
Note that this is only singular for complex momentum where we can have $\spa{a}.b=0$ without necessarily having
$\spb{a}.b=0$.    
For graviton scattering amplitudes the equivalent effective vertex is
$$
 V^{grav}(a^+,b^+, P^+) = {i  \over 360 \pi^2 } {
  {(  \spb{a}.b \spb{b}.P \spb{P}.a )  })^2
\over s_{ab} }
\equn
$$

In this article we develop techniques to determine the sub-leading pole in the amplitude.  
Double poles generally are present in most non-supersymmetric amplitudes but the starting point to study these is in the simpler purely rational amplitudes. 
At one-loop level these are
the all-plus amplitude $M_n(+,+,+,\cdots ,+)$ and the single-minus amplitude $M_n(-,+,+,\cdots, +)$.  These two amplitudes are vanishing at tree level. 
Of these the all-plus amplitude does not have double poles since the leading pole will multiply a vanishing tree amplitude.  The single-minus amplitude however
does have double poles in $s_{bc}$ with $b,c$ positive helicity legs. The leading term has 
a MHV tree amplitude as a factor.   For Yang-Mills the 
single-minus all-$n$ forms are known~\cite{Mahlon:1993si} 
and in ref.~\cite{Bern:2005hs,Vaman:2008rr} a general form for the sub-leading was presented for this specific amplitude.  The equivalent gravity amplitude has explicit forms for four~\cite{Bern:1993wt}, five and six-points~\cite{Dunbar:2010xk}
and but as yet no conjectured all-$n$ sub-leading behaviour.

 In this article we will isolate the sub-leading term which is not present in the explicit on-shell factorisation
and give expressions for this extra term.  This allows a recursive expression for the all-$n$ amplitude to be given.   The form of this is not particularly compact but it is explicit. 

As an application we examine the soft-limits of the amplitude.
Gravity amplitudes obey soft-factorisation theorems at tree level~\cite{Weinberg,White:2011yy,Cachazo:2014fwa}. Motivated by the presence of BMS symmetry~\cite{Bondi:1962px},  
these are conjectured  to also apply at loop level~\cite{Cachazo:2014fwa}. 
We confirm the result, obtained for $n\leq 6$~\cite{He:2014bga}, that one-loop
amplitudes do not obey the soft-factorisation theorems.  

\section{Single-Minus Amplitude Computation}

\def\betax{b}
\def\gammax{\gamma}

We now turn to the explicit computation of the $n$-point single-minus amplitude in gravity
$M_n(a^-,b^+,\cdots ,n^+)$\footnote{We
 use the normalisation for the full physical amplitudes
${\cal  M}^\tree=i(\kappa/2)^{n-2} M^\tree$,
\break
${\cal  M}^\oneloop=i(2\pi )^{-2} (\kappa/2)^{n} M^\oneloop$.}.

We utilise a shift on the negative helicity leg, $a$, and one of the positive helicity legs $b$,
\begin{equation}
\bar\lambda_{{a}}\to \bar\lambda_{\hat{a}} =\bar\lambda_a - z \bar\lambda_b,
\qquad , \qquad
\lambda_{{b}}\to\lambda_{\hat{b}} =\lambda_b + z \lambda_a 
\label{BCFWshiftB}
\end{equation}
We will assume, to be justified later, that under this shift $M(z)\longrightarrow 0$ as $z\longrightarrow \infty$ so that $M(0)$ 
may be determined by its residues.  This shift appears to have good asymptotic behaviour, $M\sim z^{-2}$ for $n\leq 7$,
but excites double poles.

Under the shift \eqref{BCFWshiftB} 
the factorisation channels of $M^{\rm 1-loop}(\hat a^-,\hat b^+,c^+,\cdots,n^+)$
can be arranged into terms which arise from factorisations into products of on-shell tree and loop amplitudes and an extra contribution which must
be computed separately. 

\begin{figure}[H]
\centerline{
    \begin{picture}(-50,100)(50,-25)    
     \ArrowLine(60,0)(0,0)
     \Line(0,0)(-20,-20)
     \Line(0,0)(-30,0)
     \Line(0,0)(-20,20)
     \Line(60,0)( 80,-20)
     \Line(60,0)( 90,0)
     \Line(60,0)( 80,20)
     \BCirc(0,0){8} 
     \BCirc(60,0){8} 
     \Text(60,0)[c]{$T$} 
     \Text( 0,0)[c]{$L$} 
     \Text(-25,-28)[c]{$\hat b^+$}
     \Text( 88,-28)[c]{$\hat a^-$}
     \Text( 108,12)[c]{$\{ R\}$}
     \Text(-45,12)[c]{$\{ \bar R\}$}
     \Text( 88,12)[c]{$\cdots$}
     \Text(-25,12)[c]{$\cdots$}
     \Text(12,6)[c]{$+$}
     \Text(48,6)[c]{$-$}
     \Text(30,8)[c]{$\hat P$}
    \end{picture}
    }
    \caption{Factorisations of type A.  $\{ R\}  \cup \{ \bar R\}=\{c,\cdots,n\}$ and both contain at least two elements.}  
\label{fig:typeA}\end{figure}
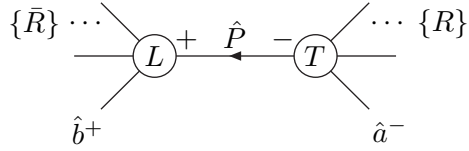

\begin{figure}[H]
\centerline{
    \begin{picture}(-50,50)(50,-50)    
     \ArrowLine(60,0)(0,0)
     \Line(0,0)(-20,-20)
     \Line(0,0)(-30,0)
     \Line(0,0)(-20,20)
     \Line(60,0)( 80,-20)
     \Line(60,0)( 80,20)
     \BCirc(0,0){8} 
     \BCirc(60,0){8} 
     \Text(60,0)[c]{$T$} 
     \Text( 0,0)[c]{$L$} 
     \Text(-25,-28)[c]{$\hat b^+$}
     \Text( 88,-28)[c]{$\hat a^-$}
    \Text( 88, 28)[c]{$\gamma+$}
     \Text(-25,12)[c]{$\cdots$}
     \Text(12,6)[c]{$+$}
     \Text(48,6)[c]{$-$}
     \Text(30,8)[c]{$\hat P$}
    \end{picture}
    }
    \caption{Factorisation of type B.}
    \label{fig:typeB}
\end{figure}
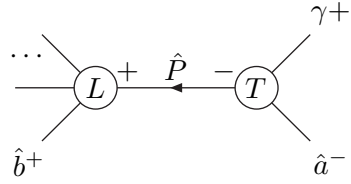

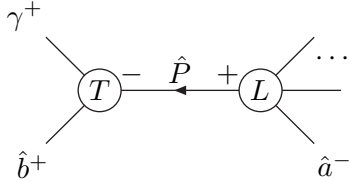
\begin{figure}[H]
\centerline{
    \begin{picture}(-50,50)(50,-50)    
     \ArrowLine(60,0)(0,0)
     \Line(0,0)(-20,-20)
     \Line(0,0)(-20,20)
     \Line(60,0)( 80,-20)
     \Line(60,0)( 90,0)
     \Line(60,0)( 80,20)
     \BCirc(0,0){8} 
     \BCirc(60,0){8} 
     \Text(60,0)[c]{$L$} 
     \Text( 0,0)[c]{$T$} 
     \Text(-25,-28)[c]{$\hat b^+$}
     \Text( 88,-28)[c]{$\hat a^-$}
    \Text( -28, 28)[c]{$\gamma^+$}
     \Text( 88,12)[c]{$\cdots$}
     \Text(12,6)[c]{$-$}
     \Text(48,6)[c]{$+$}
     \Text(30,8)[c]{$\hat P$}
    \end{picture}
    }
    \caption{Factorisation of type C.}
    \label{fig:typeC}
\end{figure}

The on-shell factorisations fall into three sets of diagrams which we label $A$,$B$ and $C$ together with the extra term labeled $\Delta_n$. 
The full amplitude is the sum of the contributions of each type:
\begin{align}
 M^{\rm 1-loop}(a^-,b^+,c^+,\cdots,n^+)= &
 A^{}_{\rm multi}(a^-,b^+,c^+,\cdots,n^+) + 
 B^{}_{\rm 3ptMHV}(a^-,b^+,c^+,\cdots,n^+)  \notag\\ & +
 C^{}_{\rm 3pt\overline{MHV}}(a^-,b^+,c^+,\cdots,n^+)+
 \Delta^{}_{n}(a^-,b^+,c^+,\cdots,n^+).
\end{align}
The first three terms may be written down straightforwardly in terms of shifted lower point amplitudes. Firstly,
\begin{equation}
A^{}_{\rm multi}(a^-,b^+,c^+,\cdots,n^+)= \sum_{\rm Part} M^{\rm tree}_{n+2-r}(\hat a^-, \hat P^-,\{R\}){1\over P^2_{a,\{R\}} }
M^{\rm 1-loop}_r(-\hat P^+,\hat b^+,\{\bar R\})\biggr\vert_{\hat P^2=0}
\end{equation}
where the sum is over all distinct partitions  of $\{c,\cdots, n\}$  into two sets $\{R\}$ and $\{\bar R\}$ which each contain at least two members.  The on-shell amplitudes are evaluated at the value of $z$ such that $\hat P^2=0$.
When the set $\{R\}$ contains a single element we have the second type of term, 
\begin{equation}
B^{}_{\rm 3ptMHV}(a^-,b^+,c^+,\cdots,n^+)= \hskip-5pt\sum_{\gamma\in\{c,\cdots, n\}}\hskip-10pt 
M^{\rm tree}_3(\hat a^-, \hat P^-,\gamma^+){1\over s_{a\gamma} }M^{\rm 1-loop}_{n-1}(-\hat P^+,\hat b^+,c^+,\cdots,\slash\hskip-6pt\gamma,\cdots,n^+)\biggr\vert_{[\hat a\gamma]=0}
\end{equation}
where $\slash\hskip-6pt\gamma$ denotes that leg $\gamma$ is omitted from the argument list. 
When the set $\{\bar R\}$ contains a single element there are also 
the factorisations, 
\begin{equation}
 C_{\rm 3pt\overline{MHV}}(a^-,b^+,c^+,\cdots,n^+) = \hskip-5pt\sum_{\gamma\in\{c,\cdots, n\}}\hskip-10pt 
 M^{\rm tree}_{3}(-\hat P^-, \hat b^+,\gamma^+)
{1\over s_{b\gamma} }M^{\rm 1-loop}_{n-1}
 (\hat a^-,\hat P^+,c^+,\cdots,\slash\hskip-6pt\gamma,\cdots,n^+)\biggr\vert_{\la\hat b\gamma\ra=0}
\end{equation}
These factorisations are illustrated in figs~\ref{fig:typeA}-\ref{fig:typeC}.

\vskip 1.0truecm

\vskip 1.0truecm

Note that these factorisations require a knowledge of both the all-plus and single-minus lower point amplitudes. 
Only the {\it n--}1 point single-minus amplitude appears. 

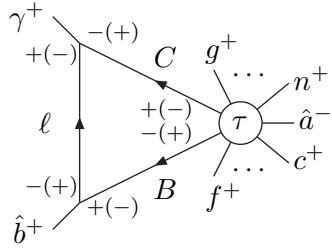
\begin{figure}[H]
\centerline{
    \begin{picture}(-50,70)(50,-50)
      \ArrowLine(0,-30)(0,30)
      \ArrowLine(60,0)(0,30) 
      \ArrowLine(60,0)(0,-30)
      \Line(0,-30)(-10,-40) 
      \Line(0, 30)(-10, 40)
      \Line(60,0)(50, 20) 
      \Line(60,0)(50,-20) 
      \Line(60,0)(78,15) 
      \Line(60,0)(80,0) 
      \Line(60,0)(78,-15)
      \BCirc(60,0){8} 
      \Text(60,0)[c]{$\tau$} 
      \Text(-13,-40)[r]{$\hat{b }^+$}
      \Text(-13, 40)[r]{$\gamma^+$}
      \Text(-13,  0)[c]{$\ell$}
      \Text(82,2)[l]{$\hat{a}^-$} 
      \Text(80, 17)[l]{$n^+$} 
      \Text(80,-13)[l]{$c^+$} 
      \Text(47, 28)[l]{$g^+$} 
      \Text(47,-28)[l]{$f^+$} 
      \Text(57, 18)[l]{$\cdots$} 
      \Text(57,-18)[l]{$\cdots$} 
      \Text(43, 4)[r]{${}^{+(-)}$} 
      \Text(43,-5)[r]{${}^{-(+)}$} 
      \Text(23,-33)[r]{${}^{+(-)}$} 
      \Text(0,-25)[r]{${}^{-(+)}$} 
      \Text(0,25)[r]{${}^{+(-)}$} 
      \Text(23,33)[r]{${}^{-(+)}$} 
      \Text(32,-25)[c]{$B$} 
      \Text(32, 25)[c]{$C$}
    \end{picture}
    }
\caption{Non-factorising contribution arising along with  the double poles.}
\label{fig:generalintegrand}
\end{figure}

The remaining contribution can be determined~\cite{Dunbar:2010xk,Alston:2012xd} by considering diagrams of the form of
fig~\ref{fig:generalintegrand}.   
We use an axial gauge formalism~\cite{Schwinn:2005pi,Kosower:1989xy} in which helicity labels can be used for internal lines and 
off-shell internal legs in the vertices are nullified using a reference spinor: given a reference null momentum $\eta$, any off-shell leg with momentum $K$ 
can be {\it nullified} using 
$$
K^\flat= K- {K^2\over [\eta| K | \eta \rangle }  \eta 
\equn$$
which gives spinors
$$
\lambda_K= \alpha K|\eta]  \;  , \;\   \bar\lambda_K= \alpha^{-1} { K|\eta\rangle \over [\eta| K | \eta \rangle } \; .
\equn$$  
Such a formalism 
has not been completely specified for gravity, however we only need a few simple properties of the three point vertex,
specifically
$$
V_3^{grav}(\alpha^+,\beta^+, \gamma^-)
=( V_3^{Y-M}(\alpha^+,\beta^+, \gamma^-) )^2
=\left(  \frac{ \spb{\alpha}.{\beta} \spa{\gamma}.\eta^2   }{ 
\spa{\alpha}.\eta \spa{\beta}.\eta  } 
\right)^2
\; . 
\equn
$$
We choose $\lambda_\eta=\lambda_a$ and $\bar\lambda_\eta=\bar\lambda_b$ to simplify
the computation. 

\vskip 1.0truecm
In \figref{fig:generalintegrand}
\begin{equation}
B=\ell+\hat b \;\;\  C=-\ell+\gamma
\end{equation}
so that
\begin{equation}
B+C=  \hat b +\gamma 
\end{equation}
and so although $B$ and $C$ are loop momentum dependent $B+C$ is not and
$s_{BC}=s_{\hat b \gamma}$. 

The diagram
shown in \figref{fig:generalintegrand} gives a contribution
\begin{align}
\int d^{D}\ell 
\Biggl(\frac{ [ b| \ell| a\ra [\gamma|\ell|a\ra }{ \la  b a \ra \la \gamma a \ra }{\la Ca\ra^2 \over \la Ba\ra^2}\Biggr)^2
\frac{ 1}{\ell^2 B^2 C^2} \,\tau(B^-,C^+,g^+\cdots n^+,\hat a^-,c^+,\cdots,f^+)
\label{basictauint}
\end{align}
which  involves 
the current $\tau(B^-,C^+,g^+\cdots n^+,\hat a^-,c^+,\cdots,f^+)$ with two off-shell  legs $B$ and $C$. 
Fortunately, as we are only interested in the residues generated by these diagrams we do not need $\tau$ exactly, only its leading and sub-leading behaviour at the pole. 
The loop integration introduces a factor of $s_{\hat b\gamma}$ for each $s_{BC}$, $B^2$ or $C^2$ appearing 
in $\tau$ , thus we can consider the expansion of $\tau$ in these 
quantities.  To see how this arises, we perform a standard Feynman parametrisation of the loop propagators:
\begin{align}
{1\over \ell^2 B^2 C^2} &\to 2\int_0^1 dx_1 dx_2 dx_3  \delta(x_1+x_2+x_3 -1){
1\over \bigl[\bigl(\ell+x_2 \hat b-x_3 \gamma\bigr)^2 + x_2 x_3 s_{\hat b \gamma}\bigr]^3}
\end{align}
and make the change of variables:
\begin{align}
 \ell&=p-x_2 \hat b+x_3 \gamma 
 \notag \\
 B&=p+(1-x_2) \hat b+x_3 \gamma 
 \notag \\
 C&=-p -x_2 \hat b- (1-x_3) \gamma 
\end{align}
leading to
\begin{align}
{1\over \ell^2 B^2 C^2}=
2\int_0^1 dx_1 dx_2 dx_3  \delta(x_1+x_2+x_3 -1){1\over \bigl[p^2 + x_2 x_3 
s_{\hat b\gamma}\bigr]^3}
\end{align}
In the rest of the integrand the change of variables replaces $s_{BC}$, $B^2$ and $C^2$ with bilinear combinations of the loop momentum, $\hat b$ and $\gamma$.
In dimensional regularization the standard tensor integral is,
\begin{align}
\int{ d^{D} \ell \over (2\pi)^{2w} }&{\ell^{a_1} \cdots \ell^{a_n} 
\over (\ell^2+2Q\cdot \ell +M^2)^A}
=
{ (-1)^n \over (4\pi)^w } {\Gamma(A-D/2)\over \Gamma(A)} 
\Biggl[{Q^{a_1} \cdots Q^{a_n}\over \bigl( Q^2-M^2\bigr)^{A-D/2} } 
+{\cal O}\biggl({1\over \bigl( Q^2-M^2\bigr)^{A-D/2-1} }\biggr) \Biggr]. 
\label{tensorint}
\end{align}
In this case $Q=0$ and $M^2=s_{\hat b\gamma}$, thus the leading term has a factor of $s_{\hat b\gamma}^{-1}$ and
any product of two loop momentum factors in the integrand will generate a factor of $s_{\hat b\gamma}$. 
With this in mind, each factor
of $s_{BC}$, $B^2$ or $C^2$ in the numerator of $\tau$ will ultimately introduce a factor of $s_{\hat b\gamma}$ which reduces the order of any pole.

\section{Gravity Currents}

In order to determine \eqref{basictauint} we would ideally use the full gravity  MHV current with two massive legs,
\begin{equation}
 \tau^{\rm MHV}_{\rm grav}(a^-,B^-,C^+,d^+, \cdots,n^+)
\end{equation}
where $B$ and $C$ are non-null, however this is not available.
Fortunately  we only need the leading and sub-leading terms in the current around $B^2=C^2=s_{BC}=0$.
In this region we may use an approximation to the current which
arises from the Kawai Lewellen and Tye (KLT) relations for on-shell amplitudes~\cite{Kawai:1985xq}
which we continue (sufficiently) off-shell.  While the  KLT relations only hold exactly on-shell,  
it turns out that we can use them to compute the pole we require from YM currents. 
The justification for this is explicitly given in the appendix and by the fact it generates a consistent amplitude. 

The explicit form for the KLT relations used is~\cite{Bern:1998sv}
\begin{align}
M_n(B, &C, . . . , n, \hat a) = i(−1)^{n+1}\Biggl[ 
A^{\rm YM}_n(B, C, . . . , n, \hat a)\times
\notag \\
&\sum_{ {\cal P}_1,{\cal P}_2}
f(i_1 , . . . , i_k ; B)\bar f(j_1 , . . . , j_{k'} ) 
A^{\rm YM}_n(i_1 , . . . , i_k , B, n, j_1 , . . . , j_{k'} ,\hat a) \Biggr] 
\notag \\
&+ {\cal P}_{KLT}(C, d, . . . , n-1),
\label{NptKLT}
\end{align}
where ${\cal P}_{KLT}$ represents a permutation over the {\it n--}3 legs $\{C,\ldots,n-1\}$. These {\it n--}3 legs are split into
two subsets; for even $n$, $\{C, . . . , n/2\}$ and $\{n/2 + 1, \ldots , n-1\}$. The
${\cal P}_1$ are the permutations  $\{i_1 , . . . , i_k\}$ of $\{C, . . . , n/2\}$, ${\cal P}_2$ 
are the permutations  $\{j_1 , \ldots , j_{k'} \}$ of $ \{n/2 + 1, \ldots , n-1\}$, $k = n/2-1$ and $ k '= n/2-2$. 
For $n$ odd, $n/2$ is replaced by $(n+1)/2$.  The KLT expression is extremely useful, however the number of terms grows rapidly with $n$. For $n$ even the
total number of terms is
$$
(n-3)! \times ( {n\over 2}-1)!  \times ( {n\over 2}-2)! 
\equn
$$

The functions $f$ and $\bar f$ are given by 
\begin{align}
 f(i_1 , . . . , i_k ; B) = & s_{B i_k}\prod_{m=1}^{k-1}\biggl(s_{B i_m } +\sum_{p=m+1}^{k}g(i_m , i_p )\biggr)
 \notag \\
 \bar f (j_1 , . . . , j_{k'} ) = &s_{j_1 n}\prod_{m=2}^{k'}\biggl(s_{j_m n} +\sum_{p=1}^{m-1}g(j_p , j_m )\biggr)
\end{align}
where
\begin{equation}
 g(i, j) = \begin{cases}
s_{ij} & {\rm if} \;\; i > j \\
                  0   & {\rm otherwise} 
\end{cases}
\label{gdefn}
\end{equation}
The inequality above is interpreted in terms of the ordering of the legs in the original gravity amplitude, i.e. $i>j$ means legs $i$ is to the right
of leg $j$ in the argument list of $M_n$ in \eqref{NptKLT}. This definition is applied to the term of ${\cal P}_{KLT}$ shown explicitly in
\eqref{NptKLT} and the permutation ${\cal P}_{KLT}$ is applied to the resulting function.   

The particular form of the KLT relation has been chosen to help us isolate the double pole term and allow 
us to compute the subleading term from this. 
In appendix~\ref{AppendixAA} it is shown explicitly  that  a quantity  $\tau^{g}_n$
that captures the leading and sub-leading behaviour of $\tau^{MHV}_{\rm grav}$ can be obtained from the KLT relations \eqref{NptKLT} 
by replacing the Yang-Mills amplitudes by Yang-Mills currents. When legs $B$ and $C$ are adjacent the Yang-Mills current is
\begin{align}
\tau^{\rm YM}_n(\hat a^- ,b^+,& \ldots, f^+, B^-,C^+,g^+,\ldots,n^+)
={\la \B a\ra^2\over \la \C a\ra^2}
{1\over \la ab\ra \cdots \la ef\ra}
 {1\over \la gh\ra \cdots \la na\ra}
\notag \\
\times
\Biggl( &
{ \spa{a}.B \spa{a}.C \spa{a}.g \spa{f}.a \over \spa{f}.B \spa{C}.g \spa{g}.f }
+
{ \spa{a}.B \spa{a}.f^2  \over \spa{B}.f \spa{g}.f } {[\eta|C|a\ra\over [\eta|B+C|f\ra}
+
{ \spa{a}.C \spa{a}.g^2  \over \spa{C}.g \spa{g}.f } {[\eta|B|a\ra\over [\eta|B+C|g\ra}
\notag \\
&-
{ \la a|BC|a\ra  \over s_{BC} } {[\eta|B+C|a\ra^2\over [\eta|B+C|f\ra[\eta|B+C|g\ra}
\Biggr)
+{\cal O}(\B^2) +{\cal O}(\C^2)  
\label{tauexpA}
\end{align}
For  the special case $\tau^{\rm YM}_n(\hat a^- , B^-,C^+,g^+,\ldots,n^+)$ we have
\begin{align}
\tau^{\rm YM}_n(\hat a^- , B^- & ,C^+,g^+,\ldots,n^+)
={\la \B a\ra^2\over \la \C a\ra^2}
 {1\over \la gh\ra \cdots \la na\ra}
\notag \\
\times \Biggl( &
{ \spa{a}.C \spa{g}.a  \over \spa{C}.g  } {[\eta|B|a\ra\over [\eta|B+C|g\ra}
 -
{ \la a|BC|a\ra  \over s_{BC} } {[\eta|B+C|a\ra\over [\eta|B+C|g\ra}
\Biggr)
+{\cal O}(\B^2) +{\cal O}(\C^2)  
\label{tauexpB}
\end{align}
When $B$ and $C$ are separated the current reduces to the Parke-Taylor form.

As discussed above, one factor of $s_{\hat b \gamma}^{-1}$ arises from the integration, so double poles in the amplitude arise when single powers of
$s_{BC}^{-1}$ occur in $\tau^{g}_n$. While there appear to be double poles in $\tau^{g}_n$ when ${\cal P}_{KLT}$ leaves $C$ adjacent to $B$ 
in the first Yang-Mills factor and $i_k=C$ in the second, there  is a factor of  $s_{BC}=s_{\hat b\gamma}$  in $f$ which
lowers the order of the pole in these cases.
Furthermore, in these terms with legs $B$ and $C$ adjacent in both Yang-Mills amplitudes the loop momentum dependent factors of $f$ have a restricted form. 
Consider first the term in ${\cal P}_{KLT}$ explicitly shown in \eqref{NptKLT} when  $i_k=C$. In this case $\bar f$ has no loop
momentum dependence. Other than $B$ (which is not involved in the permutations), $C$ is the left most leg
in the argument list of $\tau^{g}_n$, therefore when $i_k=C$, the final term in the sum: $\sum_{p=m+1}^{k}g(i_m , i_p )$ is always $g(i_m,C)=s_{i_mC}$.
All of the factors in the product part of $f$ therefore contain:
\begin{align}
s_{B i_m}+s_{C i_m} = B^2 +C^2 +[i_m|B+C|i_m\ra = B^2 +C^2 +[i_m|\hat b+\gamma|i_m\ra.
\label{fbfform}
\end{align}
As ${\cal P}_{KLT}$ doesn't involve $B$, any of these permutations that leave $C$ alone  leave both $B$ and $C$ untouched and the arguments above
are unaffected. Those elements of  ${\cal P}_{KLT}$ that move $C$ around separate $B$ and $C$ in the first Yang-Mills factor and hence are not in this class. 

An $s_{BC}$ factor also arises when legs $B$ and $C$ are adjacent in just one of the Yang-Mills factors, however the 
loop momentum dependence of the $f\bar f$ factors in these is not always simple. 

Crucially the only pole in $\tau^{g}_n$ occurs in the terms with $B$ and $C$ adjacent in both Yang-Mills factors, hence
$\tau^{g}_n$ only needs to be determined beyond leading order when $B$ and $C$ are adjacent in both Yang-Mills factors.

The KLT based expression for $\tau^g_n$ can now be used to evaluate the non-factorising contributions to the amplitude: 
$\Delta_n$. 
Using the explicit form of the KLT relations \eqref{NptKLT}, 
the non-factorising diagram depicted in \figref{fig:generalintegrand}  gives,

\begin{align}
\Delta_n^{pre-res}= &i(-1)^{n+1}
\int d^{D}\ell 
\Biggl(\frac{ [ b| \ell| a\ra [\gamma|\ell|a\ra }{ \la  b a \ra \la \gamma a \ra }{\la Ca\ra^2 \over \la Ba\ra^2}\Biggr)^2
\frac{ 1}{\ell^2 B^2 C^2} \notag \\ &
\times\Biggl[ 
\tau^{\rm YM}_n(B, C, . . . , n, a)
\sum_{ {\cal P}_1,{\cal P}_2}
f(i_1 , . . . , i_k ; B)\bar f(j_1 , . . . , j_{k'} ) 
\tau^{\rm YM}_n(i_1 , . . . , i_k , B, n, j_1 , . . . , j_{k'} , a)\Biggr]
\notag \\
&+ {\cal P}_{KLT}(C, d, . . . , n-1)
\label{eq:fullloopintegralB}
\end{align}
where $\Delta_n^{\rm pre-res}$ denotes that the residue has yet to be extracted.
Within the summation there are four distinct cases that are treated separately. These are distinguished by legs $B$ and $C$ being adjacent or separated in the two Yang-Mills currents. We denote the
contributions to the final amplitude from terms of each type by $\Delta_{SS}$, 
$\Delta_{AS}$, 
$\Delta_{SA}$ and $\Delta_{AA}$.

For terms where legs $B$ and $C$ are separated in both Yang-Mills currents, the Yang-Mills currents are finite at $\la \hat b \gamma\ra=0$ 
and they are simply the
Parke-Taylor amplitudes.
The factor
\be
\Biggl({\la Ca\ra^2 \over \la Ba\ra^2}\Biggr)^2\tau^{\rm YM}_n(B, \cdots,C, \cdots , n, a)\tau^{\rm YM}_n(i_1 , . . . , i_k , B, n, j_1 , . . . , j_{k'} , a)
\ee
then carries no net spinor weight in $C$ or $B$ and does not contain either $\bar\lambda_B$ or $\bar\lambda_C$.
Using
\be
{\la C X\ra \over \la C Y\ra}={\la C X\ra \la \gamma a\ra \over \la C Y\ra\la \gamma a\ra}=
{\la C a\ra \la \gamma X\ra +\la C \gamma\ra \la X a\ra    \over
 \la C a\ra \la \gamma Y\ra +\la C \gamma\ra \la Y a\ra}
 ={ \la \gamma X\ra  \over  \la \gamma Y\ra} +{\cal O}(\la C \gamma\ra) 
\ee
$\lambda_C$ can be replaced by $\lambda_\gamma$ in this factor. $\lambda_B$ could similarly be replaced by $\lambda_{\hat b}$. 
As there is only a single pole in these contributions, the residue simply involves this factor evaluated at 
$\la\hat b \gamma\ra=0$. Since
this factor contains no net
spinor weight in $B$, in the residue $\lambda_B$ is ultimately replaced by $\lambda_\gamma$.

As discussed above, the only contributions to the residue come from scalar integrals, therefore within the $f$ and $\bar f$ factors terms involving $B$ and $C$ are substituted as:
\begin{align}
 s_{CX}\to (1-x_3)s_{\gamma X}+x_2     s_{\hat b  X} \qquad 
 s_{BX}\to    x_3 s_{\gamma X}+(1-x_2) s_{\hat b X} \qquad 
\end{align}
The residue again involves these quantities evaluated on the pole where
\begin{align}
 s_{CX}\to (1-x_3)s_{\gamma X}+x_2     [X| b|a\ra{\spa{\gamma}.X\over \spa{\gamma}.a} \qquad 
 s_{BX}\to    x_3 s_{\gamma X}+(1-x_2) [X| b|a\ra{\spa{\gamma}.X\over \spa{\gamma}.a}
 \label{BCsubs}
\end{align}
After making these substitutions, the $f\bar f$ factor in each term in this class can be expanded:
\begin{equation}
 x_2x_3[ f \bar f]_i := \sum_r \sum_v H^{SS}_{i:rv}x_2^r x_3^v
\end{equation}
where $i$ labels the different permutations in this class.
The contribution of term $i$ to $\Delta_{SS}$ is then
\begin{align}
 \Delta_{SS:i}= i(-1)^{n+1}&{\spb{b}.{\gamma}^4\over 
 s_{b \gamma}}\sum_r \sum_v H^{SS}_{i:rv} { r! v!\over (r+v+2)!} \notag \\ 
&\times\Biggl[ 
A^{\rm YM}_n(\gamma^-,\cdots, \gamma^+, \cdots , n, a)
A^{\rm YM}_n(i_1 , \cdots, \gamma^+, \cdots, i_k , \gamma^-, n, j_1 , . . . , j_{k'} , a)\Biggr]
\end{align}
where the $\gamma$'s in the argument lists of the Yang-Mills amplitudes appear where $B$ and $C$ appeared originally. 

As discussed above, when legs $B$ and $C$ are adjacent in either of the Yang-Mills currents in \eqref{eq:fullloopintegralB}, $f\bar f$ always contains a factor of $s_{BC}$. If legs $B$ and $C$ are adjacent
in just one of the currents, there is a single pole overall,  only the singular part of the current with $B$ and $C$ adjacent contributes and the Parke-Taylor form can be used for the other. 
Again we are dealing with a simple pole, so in the residue all shifted quantities are evaluated on the pole.

If legs $B$ and $C$ are adjacent in only the first 
Yang-Mills current, $\tau^{\rm YM}_n(\hat a , B, C,g, . . . , n)$, the Parke-Taylor amplitude is used for the second current with the same substitutions
as in the separated-separated case. The Feynman parameter integrations can also be performed  in a similar fashion: after making the substitutions 
\eqref{BCsubs}, $f\bar f$ is again expanded for each term in the class:
\begin{equation}
 x_2x_3(x_3+x_2-1) [f \bar f]_i := \sum_r \sum_v H^{AS}_{i:rv}x_2^r x_3^v
\end{equation}

To help track the leg orderings, the string of $\la xy\ra$ factors in the denominator of \eqref{tauexpA} is denoted 
\begin{align}
F^{\rm YM}_n(a^- ,b^+,& \ldots, f^+, \slash\hskip-7pt B^-,\slash\hskip-7pt C^+,g^+,\ldots,n^+)
={1\over \la ab\ra \cdots \la ef\ra}{1\over \la gh\ra \cdots \la na\ra}
\end{align}
where $ \slash\hskip-7pt B^-$ and $\slash \hskip-7pt C^+$ denote the positions of these legs in the original current and the slashes highlight that the function does not depend on the values of $B$ and $C$.
$g$ denotes the leg following $C$ in the first Yang-Mills current. The contribution of each term in this class to $\Delta_{AS}$ is 
\begin{align}
 \Delta_{AS:i}= i&(-1)^{n+1}{[b\gamma]^4 \la a|b\gamma|a\ra\over s_{b\gamma}}\sum_r \sum_v H^{AS}_{i:rv} { r! v!\over (r+v+2)!} \notag \\ 
&\times\Biggl[ 
{\spa{\gamma}.a\over \spa{\gamma}.g}F^{\rm YM}_n(a, \slash \hskip-7pt B,\slash \hskip-7pt C,g, \cdots , n)
A^{\rm YM}_n(i_1 , \cdots, \gamma^+, \cdots, i_k , \gamma^-, n, j_1 , . . . , j_{k'} , a)\Biggr]
\end{align}

When legs $B$ and $C$ are adjacent in only the second Yang-Mills current,

\noindent{$\tau^{\rm YM}_n(i_1 , . . . g, C , B, n, j_1 , . . . , j_{k'} , \hat a)$,} the analogous definitions are
\begin{equation}
 x_2x_3(x_3+x_2-1) [f \bar f]_i := \sum_r \sum_v H^{i:SA}_{rv}x_2^r x_3^v
\end{equation}
and leg $g$ is the leg appearing before $C$ in the second  Yang-Mills current. The contribution of  each term in this class  to $\Delta_{SA}$ is 
\begin{align}
 \Delta_{SA:i}= i&(-1)^{n+1}{[b\gamma]^4 \la a|b\gamma|a\ra\over s_{b\gamma}}\sum_r \sum_v H^{i:SA}_{rv} { r! v!\over (r+v+2)!} \notag \\ 
&\times\Biggl[ 
A^{\rm YM}_n(a, \gamma^-,\cdots,\gamma^+,\cdots , n)   
{\spa{\gamma}.a^2\over \spa{\gamma}.g\spa{\gamma}.n}F^{\rm YM}_n(i_1 ,\cdots, g,\slash \hskip-7pt C,\slash \hskip-7pt B, n, j_1 , \cdots, j_{k'} , a)\Biggr]
\end{align}

Finally there are contributions from terms where legs $B$ and $C$ are adjacent in both Yang-Mills currents. The form (\ref{tauexpA}) is used for both currents. 
As  the $f\bar f$ piece of these terms always have a factor of $s_{b\gamma}$ there is no pole in the terms involving the product of the regular parts of the two currents,
a single pole in the terms where the singular part of one current multiplies the regular piece of the other (we denote these contributions $\Delta_{A_rA_s}$ and $\Delta_{A_sA_r}$)
and a double pole in the product of the two singular parts (we denote these contributions $\Delta_{A_sA_s}$). 

Note that, in this particular case,  the loop momentum dependence of $f\bar f$ has the very restricted form given in \eqref{fbfform}.
The pieces involving $B^2$ and $C^2$ cancel the corresponding propagator giving massless bubbles which are discarded in the usual dimensional 
regularisation prescription. The surviving piece of $f\bar f$ has no loop momentum dependence
and thus no $x$ dependence.   Although not loop momentum dependent, we must expand $f\bar f$ about the pole. This corresponds to substituting
\begin{equation}
s_{CX} \longrightarrow s_{\gamma X}
\;\;
s_{BX} \longrightarrow   [X|b | a \ra { \spa{\gamma}.X \over \spa{\gamma}.a} +\Delta z 
[b|X|a\rangle
\end{equation}
Then
\begin{equation}
{[f\bar f ] \over s_{BC} }
= [f\bar f]_0 +\Delta z [f\bar f]_1 +O(\Delta z^2)
\end{equation}
Schematically expanding the Yang-Mills currents and $f\bar f$ factors in terms of $s_{b\gamma}$, the adjacent-adjacent contribution has the form
$$
\left(  { \tau_{-1} \over s_{b\gamma} } + \tau_0 \right) s_{b\gamma} \left( [f\bar f]_0
    +s_{b\gamma}  [f\bar f]_1 \right) \left( 
   { \tau_{-1} \over s_{b\gamma} } + \tau_0   \right)
\equn
$$
which gives the  pole term 
$$
{1\over s_{b\gamma} }
\tau_{-1} \times [ f\bar f ]_0 \times \tau_{-1}
\equn
$$
along with finite pieces
$$
 \tau_{-1} \times [f\bar f]_1 \times \tau_{-1} 
+\tau_{-1} \times [f\bar f]_0 \times \tau_0 
+\tau_0    \times [f\bar f]_0 \times \tau_{-1}
\equn
$$
Extracting a common factor of
\begin{eqnarray}
C_0 &=& -{ \spb{b}.{\gamma}^4 \la a | b \gamma | a \ra  \over 360
  s_{b\gamma} } \times
F_n[ a,\slash \hskip-7pt B,\slash \hskip-7pt C,g,\cdots ,n)] 
\times F_n[i_1,\cdots,f,\slash \hskip-7pt C,\slash \hskip-7pt B,n,j_1,\cdots,j_k,a ] 
\notag \\
&=& -{ \spb{b}.{\gamma}^4 \la a | b \gamma | a \ra  \spa{a}.\gamma \spa{\gamma}.g \spa{f}.\gamma \spa{\gamma}.n \over 360 \spa{a}.\gamma^8 
  s_{b\gamma} } 
\notag
\\
& & \times
A_{n-1} [ a^-,\gamma^-,g,\cdots ,n)] \times A_{n-1}[i_1,\cdots,f,\gamma^-,n,j_1,\cdots,j_k,a^- ] 
\end{eqnarray}
the adjacent-adjacent contribution is
$$
\Delta_{AA:i}=C_0 \times \left(  { J_{r,r} \over s_{b\gamma}}
  [f\bar f]_0+J^1_{r,r}[f\bar f]_1   +   (  J_{r,s}+J_{s,r} )  [f\bar f]_0  
\right)
\equn
$$
where 
\begin{eqnarray}
J_{r,s} &=&  { \spa{\gamma}.a^4  \spa{a}.g \over \spa{\gamma}.g^2
  \spa{\gamma}.f \spa{\gamma}.n }
\\
J_{s,r} &=&  \left( 
{ \spa{\gamma}.a^3  \over \spa{\gamma}.g\spa{f}.n }
\right) \times \left( 3 { \spa{a}.f \spa{n}.a \over \spa{n}.\gamma
    \spa{\gamma}.f }-2 {\spa{a}.n^2 \over \spa{n}.\gamma^2 }
- {\spa{a}.f^2 \over \spa{f}.\gamma^2 }
\right) 
\\
J_{r,r} &=&   { \spa{\gamma}.a^3 \la a | b \gamma | a \ra \over \spa{\gamma}.{g} \spa{\gamma}.n
\spa{\gamma}.f }
\\
J_{r,r}^1 &=&   { \spa{\gamma}.a^2 \la a | b \gamma | a \ra\over \spb{b}.{\gamma} \spa{\gamma}.{g} \spa{\gamma}.n
\spa{\gamma}.f }
\end{eqnarray}

The full non-factorising contribution is then
\begin{equation}
\Delta_n(a^-,b^+,c^+,\cdots,n^+)=\sum_i 
\Delta_{SS:i}+\sum_i \Delta_{AS:i}
+\sum_i \Delta_{SA:i}+\sum_i \Delta_{AA:i}
 \end{equation}
where each sum is over all terms in the relevant class in \eqref{eq:fullloopintegralB}.
 
We have used these expressions to generate the single-minus amplitudes for $n \leq 8$: our expressions have the correct symmetries and collinear limits
as detailed in Appendix~\ref{MathematicaAppendix}.

\section{Soft Limits}

Graviton scattering amplitudes are singular as a leg becomes
soft. Weinberg~\cite{Weinberg} many years ago presented the leading
soft limit. 
If we parameterise the momentum of the $n$-th leg as $k_n^\mu = t \times k_s^\mu$ then  
in the
limit $t\longrightarrow 0$ the singularity in the $n$-point amplitude is
$$
M_n \longrightarrow {1\over t} \times S^{(0)} \times M_{n-1} +O(t^0)
\equn
\label{softone}
$$
where $M_{n-1}$ is the {\it n--}1 point amplitude.    The soft-factor
$S^{(0)}$ is
universal and  Weinberg showed that \eqref{softone} does not receive corrections in loop amplitudes.
 
\def\pol{\eps}
\def\sand#1.#2.#3{%
\left\langle\smash{#1}{\vphantom1}^{-}\right|{#2}%
\left|\smash{#3}{\vphantom1}^{-}\right\rangle}
\def\sandp#1.#2.#3{%
\left\langle\smash{#1}{\vphantom1}^{-}\right|{#2}
\left|\smash{#3}{\vphantom1}^{+}\right\rangle}
\def\sandpp#1.#2.#3{%
\left\langle\smash{#1}{\vphantom1}^{+}\right|{#2}%
\left|\smash{#3}{\vphantom1}^{+}\right\rangle}
\def\sandmm#1.#2.#3{%
\left\langle\smash{#1}{\vphantom1}^{-}\right|{#2}%
\left|\smash{#3}{\vphantom1}^{-}\right\rangle}

Recently it has also been proposed \cite{White:2011yy,Cachazo:2014fwa} that the sub-leading and sub-sub leading terms are also universal.  This
can be best exposed, when a positive helicity leg becomes soft,  by
setting
$$
\lambda_n = t \times \lambda_s , \;\;\; 
\bar\lambda_n =  \bar\lambda_s ,
\equn\label{BasicSoft}
$$
In the $t\longrightarrow 0$ limit the amplitude now has $t^{-3}$ singularities.
Effectively the choice \eqref{BasicSoft} introduces a $t^{-2}$ factor into the
polarisation tensor $\pol_{\mu\nu}^+$ and a $t^{+2}$ factor into $\pol_{\mu\nu}^-$.
This is because the polarisation tensors for gravitons are a product of
gluonic polarisations vectors, $\pol_{\mu\nu} =\pol_\mu\pol'_\nu$, and  the gluonic polarisation vectors for a leg with momentum $k$ using
reference momentum $q$ are 
$$
\pol^{+}_{\mu} (k;q) =
{ \la q | \gamma^\mu |k ]
\over  \sqrt2 \spa{q}.k} \hskip 2.0 cm
\pol^{-}_{\mu} (k;q) =
{ [ q |\gamma^\mu | k \ra 
\over \sqrt{2} \spb{k}.q}
\equn\label{SpinorHelicity}
$$
Thus the choice \eqref{BasicSoft} engineers a $t^{-1}$
singularity in $\pol^+_\mu(k; q)$. 

At tree level the soft behaviour is~\cite{Cachazo:2014fwa}
$$
\Mtree_n 
=(  {1\over t^3} S^{(0)} +{1\over t^2} S^{(1)} +{1\over t
} S^{(2)} ) \Mtree_{n-1} +O(t^0)
\equn
$$ 
Where, for a positive helicity-leg becoming soft~\cite{Bern:2014oka,Broedel:2014fsa} 

\begin{eqnarray}
S^{(0)}&=&  -\sum_{i=1}^{n-1} { \spb{s}. i  \spa{i}.{\alpha} \spa{i}.\beta\over
  \spa{s}. i \spa{s}.{\alpha} \spa{s}.\beta }
\\
S^{(1)} &=&  -\frac{1}{2} \sum_{i=1}^{n-1}    { \spb{s}.i \over \spa{s}.i }  
\left(   { \spa{i}.\alpha \over \spa{s}.\alpha }+{ \spa{i}.\beta
    \over \spa{s}.\beta } 
\right)
\bar\lambda_s^{\dot a} { \partial \over \partial \bar\lambda_i^{\dot a} }
\\
S^{(2)} &=& \frac{1}{2} \sum_{i=1}^{n-1} { \spb{i}.s \over \spa{i}.s }
\bar\lambda_s^{\dot a} \bar\lambda_s^{\dot b}
{ \partial \over \partial \bar\lambda_i^{\dot a} } 
{ \partial \over \partial \bar\lambda_i^{\dot b} }
\end{eqnarray}

There are a few points to note regarding these expressions.   The
expressions  are  independent of the spinors $\alpha$ and
$\beta$ which arise from the
choice of reference spinors for the soft leg as in \eqref{SpinorHelicity}.  
Also, we cannot simply act with these operators on on-shell
expressions.   For an on-shell expression, the spinors satisfy
$\sum_i  \lambda^a_i \bar\lambda_i^{\dot a}  =0$
which implies that not all the variables should be treated as
independent. A solution to this is to
choose two of the $\bar\lambda$'s to eliminate and solve via\footnote{
With this prescription the momenta $k_1$ and $k_2$ are complex valued.}
$$
\bar\lambda_1 = -\sum_{i=3}^n  { \spa{2}.{i} \over \spa2.1}
\bar\lambda_i
\;\;\;
\bar\lambda_2 = -\sum_{i=3}^n  { \spa{1}.{i} \over \spa1.2 }
\bar\lambda_i
\equn
$$

The universal soft theorem has been proven to hold for tree level\cite{Schwab:2014xua,Bern:2014oka,Klose:2015xoa,Zlotnikov:2014sva,Kalousios:2014uva}
amplitudes and a bold generalisation is conjectured to remain true for loop amplitudes~\cite{Cachazo:2014fwa},
$$
\Mloop_n=(  {1\over t^3} S^{(0)} +{1\over t^2} S^{(1)} +{1\over t
} S^{(2)} ) \Mloop_{n-1} 
+O(t^0)
\equn\label{SoftLoopRidiculouslyOptimistic}
$$  
Unfortunately this does not in fact hold.   A possible extension at one-loop would be
$$
\Mloop_n=(  {1\over t^3} S^{(0)} +{1\over t^2} S^{(1)} +{1\over t
} S^{(2)} ) \Mloop_{n-1} 
+( {1\over t^2} S^{(1)}_{\oneloop} +{1\over t
} S^{(2)}_{\oneloop} ) \Mtree_{n-1} 
+O(t^0)
\equn\label{SoftLoop}
$$
where the operators $S^{(i)}_{\oneloop}$ are loop soft factors acting
upon the tree. Since the leading soft term has no loop corrections the
loop corrections must start at $t^{-2}$.  In this section we confirm and extend the evidence~\cite{Bern:2014oka,He:2014bga} for
loop amplitudes having loop corrections to the sub-sub leading terms beyond~\eqref{SoftLoop}.

The two classes of helicity amplitude which are zero at tree
level and are consequently rational at one-loop are the all-plus amplitudes, $\Mloop_n(1^+,2^+,\cdots n^+)$, and the single-minus amplitudes,
$\Mloop_n(1^-,2^+, \cdots  n^+)$.
For both of these classes the  tree amplitudes
vanish, so \eqref{SoftLoop} will reduce to \eqref{SoftLoopRidiculouslyOptimistic}.  
The two classes have different stories~\cite{He:2014bga}:  
the all-plus amplitudes have the expected behaviour \eqref{SoftLoopRidiculouslyOptimistic} whereas the
single-minus amplitudes have anomalous sub-sub-leading terms.  One difference between
the classes is the appearance of double-poles in the
single-minus amplitudes.   Previously the single-minus amplitudes were only
known for $n \leq 6$.  
Having computed  a recursive representation for the single-minus
amplitudes we have checked whether they satisfy the soft-theorem \eqref{SoftLoop}.  
For $n \leq 8$ we find

$\bullet$ \eqref{SoftLoopRidiculouslyOptimistic} is satisfied for a negative helicity leg going soft

$\bullet$ The leading and sub-leading terms in \eqref{SoftLoopRidiculouslyOptimistic} are correct for a positive helicity leg going 

\hskip 12pt soft

$\bullet$ The sub-sub-leading terms in the amplitudes {\it do not} match \eqref{SoftLoopRidiculouslyOptimistic}.

These results are in agreement with those of \cite{He:2014bga} where the soft theorems
were examined in the five and six point cases.

\section{Conclusions} 

In this article we have used augmented recursive techniques to obtain an explicit form for the non-factorising piece of the single-minus graviton scattering amplitude.   To do so we have had to identify the difficult sub-leading poles
in the recursion process.   These are endemic in non-supersymmetric one-loop computations. 

Explicit  
amplitudes have repeatedly been crucial to test and challenge various hypothesis in interacting field theory. We have generated hard to reach amplitudes 
which we hope will be useful. 
For example, we have used our expressions to confirm 
the presence of loop anomalies to the conjectured universal soft-limit of gravity amplitudes.  The explicit forms of the five, six, seven and eight-point amplitudes are available at \url{http://pyweb.swan.ac.uk/~dunbar/graviton.html}

\section{Acknowledgements}

This work was supported by STFC grant ST/L000369/1.

\appendix

\section{Using the KLT relations}
\label{AppendixAA}

The non-factorising contributions to the amplitude depend on MHV currents with two massive legs, e.g.
\begin{equation}
 \tau^{\rm MHV}_{\rm grav}(a^-,B^-,C^+,d^+, \cdots,n^+)
\end{equation}
where $B$ and $C$ are non-null. Specifically the amplitude is sensitive to both the leading and sub-leading terms in this current around $B^2=C^2=s_{BC}=0$.

Using a reference spinor, $\eta$,  a  {\it nullified}  version of any massive momentum can be defined via
\begin{equation}
 P^\flat=P-{P^2\over [\eta|P|\eta\ra}\eta 
 \,\, \to \,\,
 P(P^\flat,P^2)=P^\flat +{P^2\over [\eta|P^\flat|\eta\ra}\eta
\end{equation}
so that 
\begin{equation}
 \tau_n=\tau_n(a,B^\flat,C^\flat,d, \cdots,n,B^2,C^2).
\end{equation}
$\tau_n$ can be expanded initially around $B^2=C^2=0$, 
\begin{equation}
 \tau_n= F_1 +B^2 F_2 +C^2 F_3 + \ldots
 \qquad {\rm where} \quad
 F_i=F_i(a,B^\flat,C^\flat,d, \cdots,n).
\end{equation}
At tree level the current is not expected to have double poles, but there are single $s_{BC}$ poles. The $F_i$ can therefore be expanded about 
$s_{BC}=0$ as
\begin{equation}
F_i={G_i\over  s_{BC}} + \sum_{j=0}^{\infty} H_i^j s_{BC}^j,
\end{equation}
where the $G_i$ and $H_i^j$ are evaluated on $B^2=C^2=s_{BC} =0$.
The expansion of $\tau_n$ is then
\begin{equation}
 \tau_n= {G_1 +B^2 G_2 +C^2 G_3\over s_{BC}} +\sum_{j=0}^{\infty} H_1^j s_{BC}^j + {\cal O}(B^2)+ {\cal O}(C^2).
\end{equation}

Now consider the quantity $\tau^g_n$ which is {\it good enough} for the current purposes.
The expansion of the difference between the two quantities is,
\begin{equation}
\tau_n -\tau^g_n= {\delta G_1 +B^2\delta G_2 +C^2\delta G_3\over s_{BC}} 
+\sum_{j=0}^{\infty} \delta H_1^j s_{BC}^j + {\cal O}(B^2)+ {\cal O}(C^2).
\end{equation}

Two constraints must be imposed on $\tau^g_n$: 
\begin{align}
{\rm C.1}: & \qquad  {lim \atop s_{BC}\to 0} \biggl( s_{BC} \tau_n- s_{BC} \tau^g_n\biggr) =0
\notag \\
{\rm C.2}: & \qquad {lim \atop {B^2\to0,C^2\to 0 \atop s_{BC}\neq 0}} \biggl( \tau_n- \tau^g_n\biggr) =0
\end{align}
C.1 requires
\begin{equation}
\delta G_1 +B^2\delta G_2 +C^2\delta G_3 =0.
\end{equation}
As the $G_i$ are evaluated on $B^2=C^2=0$, this can only hold if 
$
 \delta G_1 =\delta G_2 =\delta G_3 =0.
$
While C.2 requires
 \begin{equation}
 {\delta G_1\over  s_{BC}} +\sum_{j=0}^{\infty} \delta H_1^j s_{BC}^j =0.
 \end{equation}
As the $H_1^j$ are evaluated at $s_{BC}=0$ this can only hold for any $s_{BC}$ if $\delta H_1^j=0$. Combining the constraints from C.1 and C.2,
\begin{equation}
\tau_n=\tau^g_n+ {\cal O}(B^2)+{\cal O}(C^2).
\end{equation}
Hence conditions C.1 and C.2 ensure that $\tau^g_n$ correctly encompasses the leading and sub-leading behaviour of $\tau_n$ which determine the 
non-factorising contributions to the amplitude.

$\tau^g_n$ can be obtained by a generalisation of \eqref{NptKLT} to currents:
\begin{align}
\tau^{g}_n(B, &C, . . . , n, \hat a) = i(−1)^{n+1}\Biggl[ 
\tau^{\rm YM}_n(B, C, . . . , n, \hat a)\times
\notag \\
&\sum_{ {\cal P}_1,{\cal P}_2}
f(i_1 , . . . , i_k ; B)\bar f(j_1 , . . . , j_{k'} ) 
\tau^{\rm YM}_n(i_1 , . . . , i_k , B, n, j_1 , . . . , j_{k'} ,\hat a)\Biggl]
\notag \\
&+ {\cal P}_{KLT}(C, d, . . . , n-1),
\label{NptKLTtau}
\end{align}
where the Yang-Mills currents are given by
\begin{align}
\tau^{\rm YM}_n(\hat a^- ,b^+,& \ldots, f^+, B^-,C^+,g^+,\ldots,n^+)
={\la \B a\ra^2\over \la \C a\ra^2}
{1\over \la ab\ra \cdots \la ef\ra}
 {1\over \la gh\ra \cdots \la na\ra}
\notag \\
\Biggl( &
{ \spa{a}.B \spa{a}.C \spa{a}.g \spa{f}.a \over \spa{f}.B \spa{C}.g \spa{g}.f }
+
{ \spa{a}.B \spa{a}.f^2  \over \spa{B}.f \spa{g}.f } {[\ceta|C|a\ra\over [\ceta|B+C|f\ra}
+
{ \spa{a}.C \spa{a}.g^2  \over \spa{C}.g \spa{g}.f } {[\ceta|B|a\ra\over [\ceta|B+C|g\ra}
\notag \\
&-
{ \la a|BC|a\ra  \over s_{BC} } {[\ceta|B+C|a\ra^2\over [\ceta|B+C|f\ra[\ceta|B+C|g\ra}
\Biggr)
+{\cal O}(\B^2) +{\cal O}(\C^2)  
\label{YMtauexp}
\end{align}
and the functions $f$ and $\bar f$, partitions, summations etc. are those specified for \eqref{NptKLT}.

The fact that $\tau^g_n$ as defined by \eqref{NptKLTtau} satisfies conditions C.1 and C.2 follows from the identity 
\begin{align}
& { \spa{C}.a^2 \spa{B}.a^2 \over \spa{f}.B \spa{B}.C \spa{C}.g }
+
{ \spa{B}.a \spa{C}.a  \over \spa{B}.{C} } {[\ceta|B+C|a\ra^2\over [\ceta|B+C|f\ra[\ceta|B+C|g\ra}
\notag \\
=&
{ \spa{a}.B \spa{a}.C \spa{a}.g \spa{f}.a \over \spa{f}.B \spa{C}.g \spa{g}.f }
+
{ \spa{a}.B \spa{a}.f^2  \over \spa{B}.f \spa{g}.f } {[\ceta|C|a\ra\over [\ceta|B+C|f\ra}
+
{ \spa{a}.C \spa{a}.g^2  \over \spa{C}.g \spa{g}.f } {[\ceta|B|a\ra\over [\ceta|B+C|g\ra}
\label{useful1}
\end{align}
which holds for arbitrary null momenta and when $B$ and $C$ are massive, if $\lambda_B$ etc. are understood to refer to their nullified forms.
\eqref{useful1} can be rewritten as
\begin{align}
&
{ \spa{a}.B \spa{a}.C \spa{a}.g \spa{f}.a \over \spa{f}.B \spa{C}.g \spa{g}.f }
+
{ \spa{a}.B \spa{a}.f^2  \over \spa{B}.f \spa{g}.f } {[\ceta|C|a\ra\over [\ceta|B+C|f\ra}
+
{ \spa{a}.C \spa{a}.g^2  \over \spa{C}.g \spa{g}.f } {[\ceta|B|a\ra\over [\ceta|B+C|g\ra}
\notag \\
&-
{ \la a|BC|a\ra  \over s_{BC} } {[\ceta|B+C|a\ra^2\over [\ceta|B+C|f\ra[\ceta|B+C|g\ra}
\notag \\
&= { \spa{C}.a^2 \spa{B}.a^2 \over \spa{f}.B \spa{B}.C \spa{C}.g } +{\cal O}(B^2)+{\cal O}(C^2).
\end{align}
The right hand side is precisely the factor required for \eqref{YMtauexp} to  reproduce the  Parke-Taylor amplitude in 
the $B^2,C^2\to 0$, $s_{BC}\neq0$ limit. Thus in this limit \eqref{NptKLTtau} is the standard KLT relation involving 
Yang-Mills amplitudes which reproduces the gravity amplitude as required by condition C.2.

\begin{figure}[H]
\centerline{
    \begin{picture}(150,100)(0,-10)    
     \Line(20,0)(50,20)
     \Line(20,40)(50,20)
\Line(50,20)(90,20)
\Line(80,30)(100,40)
\Line(80,10)(100,0)
\COval(80,20)(20,10)(0){Black}{Yellow} 
  \Text(25,48)[c]{$B$}
  \Text(25,-5)[c]{$C$}
    \end{picture}
    }
    \caption{Diagrams contributing to the $s_{BC}^{-1}$ poles in $\tau_n$.}  
   \label{fig:tausingfig}
   \end{figure}
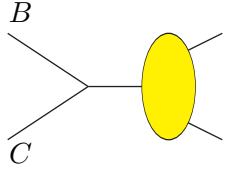

Condition C.1 involves the $s_{BC} \to 0$ limit. $\tau_n$ has $s_{BC}^{-1}$ singularities arising from diagrams of the form shown in \figref{fig:tausingfig}.

There appear to be double poles in $\tau^{g}_n$ when ${\cal P}_{KLT}$ leaves $C$ adjacent to $B$ 
in the first Yang-Mills factor and $i_k=C$ in the second. However, when $i_k=C$ there  is a factor of  $s_{BC}$  in $f$ which
lowers the order of the pole. In fact these terms  are the only ones that are singular in the $s_{BC} \to 0$ limit. 
When $B$ and $C$ are adjacent in just one of the Yang-Mills factors there is always a factor of $s_{BC}$ in $f$  which removes the singularity in the current.
If $B$ and $C$ are adjacent in the second Yang-Mills factor, $i_k=C$ and there is an explicit factor of $s_{BC}$ in $f$ as previously. 
If $B$ and $C$ are adjacent in the first Yang-Mills factor, we are considering  a term in ${\cal P}_{KLT}$ as explicitly shown in \eqref{NptKLT} where 
$C=i_q$ with $q<k$. As ${\cal P}_{KLT}$ hasn't touched $C$, we can still use the definition of $g(i,j)$ given in \eqref{gdefn}. 
In this case the product piece of $f$ contains a factor with $m=q$ which takes the form
\begin{equation}
s_{B i_q } +\sum_{p=q+1}^{k}g(i_q , i_p ) = s_{BC}+\sum_{p=q+1}^{k}g(C , i_p ). 
\end{equation}
As $C$ is the left most leg
in the argument list of $\tau^{\rm grav}_n$ and in these cases is untouched by ${\cal P}_{KLT}$, $g(C , i_p )=0$ and we again have a factor of $s_{BC}$.

In the $s_{BC} \to 0$ limit the momentum $B+C$ becomes null and setting $B+C=k$ the singular parts of the Yang-Mills current become
\begin{align}
\tau^{\rm YM:sing}_n & (\hat a^- ,b^+, \ldots, f^+, B^-,C^+,g^+,\ldots,n^+) \notag \\
&
={\la \B a\ra^2\over \la \C a\ra^2}
{1\over \la ab\ra \cdots \la ef\ra}
 {1\over \la gh\ra \cdots \la na\ra}
\Biggl( -
{ \la a|BC|a\ra  \over s_{BC} } {[\ceta|B+C|a\ra^2\over [\ceta|B+C|f\ra[\ceta|B+C|g\ra}
\Biggr)   
\notag \\
&
={\la \B a\ra^2\over \la \C a\ra^2}
{1\over \la ab\ra \cdots \la ef\ra\la fk\ra \la kg\ra \la gh\ra \cdots \la na\ra}
\Biggl( -
{ \la a|BC|a\ra  \over s_{BC} } {\la ka\ra^2}
\Biggr)
\notag \\
&
={\la \B a\ra^2\over \la \C a\ra^2}
A^{\rm YM}_{n-1}(\hat a^- ,b^+, \ldots, f^+, k^-,g^+,\ldots,n^+)
\Biggl( -
{ \la a|BC|a\ra  \over s_{BC} \la ka\ra^2}
\Biggr)  
\end{align}
The singular part of the current thus becomes
\begin{align}
\tau^{g:sing}_n= i(−1)^{n+1}\Biggl[ &
\tau^{\rm YM:sing}_n(B, C, . . . , n, \hat a)\times
\notag \\
&\sum_{ {\cal P}'_1,{\cal P}_2}
f(i_1 , . . . , i_k ; B)\bar f(j_1 , . . . , j_{k'} ) 
\tau^{\rm YM:sing}_n(i_1 , . . . , i_k , B, n, j_1 , . . . , j_{k'} ,\hat a) \Biggr]
\notag \\
&+ {\cal P}_{KLT}'(d, . . . , n-1),
\end{align}
Where ${\cal P}_{KLT}'$ and ${\cal P}'_1$ denote the subset of the original permutation sums that leave $C$ adjacent to $B$ in both Yang-Mills factors,
i.e. $C$ is excluded from ${\cal P}_{KLT}'$ and $i_k=C$.
The momentum factor $f$ is then
\begin{align}
 f(i_1 , \cdots ,i_{k-1}, C ; B) = & s_{BC}\prod_{m=1}^{k-2}\biggl(s_{B i_m } +\sum_{p=m+1}^{k}g(i_m , i_p )\biggr)
 \biggl(s_{B i_{k-1} } +g(i_{k-1} , C )\biggr)
\end{align}
As we are only interested in those permutations that leave leg $C$ unmoved, $C$ remains to the left of all possible legs $i_{k-1}$ and $g(i_{k-1} , C )=0$.
Thus 
\begin{align}
 f(i_1 , . . i_{k-1}, C ; B) = s_{BC}f(i_1 , . . i_{k-1}; B)
\end{align}
The summations and momentum factors are thus precisely those of the ({\it n--}1)-point KLT expansion.
\begin{align}
\tau^{g:sing}_n= -{\la \B a\ra^4\over \la \C a\ra^4}{ \la a|BC|a\ra^2  \over\la ka\ra^4}{1\over s_{BC} }M_{n-1}(\hat a^- ,b^+, \ldots, f^+, k^-,g^+,\ldots,n^+) &
\end{align}
Taking the three-point gravity vertex to be the square of its Yang-Mills counter part, in axial gauge with $\lambda_\ceta=\lambda_a$, we have
\begin{equation}
V(B^-,C^+,k^+)={ [Ck]^2 \la Ba\ra^4 \over \la Ca\ra^2 \la ka\ra^2 }
={\la Ba\ra^4 \over\la Ca\ra^4} {\la a|Ck|a\ra^2\over \la ka\ra^4 }
={\la Ba\ra^4 \over\la Ca\ra^4} {\la a|CB|a\ra^2\over \la ka\ra^4 }
\end{equation}
In the $B$, $C$ co-linear limit  $\tau^{g:sing}_n$ precisely reproduces the factorisation of the gravity current and hence satisfies condition C.1.

\section{Using Soft Theorems to Construct Amplitudes} 

If the soft-limit theorems hold they could be a substitute for the rather
cumbersome augmented recursion techniques.  If we take
$$
\lambda_n = t \lambda_s \;\;\;
\bar\lambda_n = \bar \lambda_s
\equn
\label{appysoft}
$$
together with a specific choice for implementing momentum
conservation,
$$
\bar\lambda_1(t)=\bar\lambda_1-t { \spa{2}.s \over \spa{2}.1 }
\bar\lambda_s \;\;\;
\bar\lambda_2(t)=\bar\lambda_2-t { \spa{1}.s \over \spa{1}.2 }
\bar\lambda_s
\equn
$$
and treat $t$ as  a complex parameter, then we have the complexified
amplitude $M_n(t)$.  We would like to construct the $n$-point
amplitude $M_n(t=t_0)$ recursively from $M_{n-1}$.  This is possible by
evaluating
$$
\int dt  { M_n(t) \over t-t_0 } \, .
\equn
$$
{\it Providing} $M_n(t) \longrightarrow 0$ as
$t\longrightarrow\infty$
we have
$$
M_n(t_0) = -\sum_i {\rm Residue} [  {M_n(t) \over t-t_0} , t_i]
\equn
$$
At values of $t$ (say $t_i$)  where $M_n(t)$ has simple poles
$$
{\rm Residue} [ { M_n(t) \over t-t_0} ,t_i] 
= {1\over t_i-t_0 } {\rm Residue}[ M_n(t_i) , t_i]
\equn
$$
which can be determined from the factorisation theorems as usual.  
When $M_n(t)$ has a non-simple pole, the residue of $M_n(t)/(t-t_0)$ in
general involves all the singular terms. For example if $M_n(t)$ has a
multiple singularity at $t=0$ and the expansion about $t=0$ is
$$
M_n(t)= {a_{-N} \over t^N} +
{a_{-N+1} \over t^{N-1}} \cdots { a_{-1} \over t}+\hbox{\rm non-singular}
\equn
$$
then
$$
{\rm Residue} [ { M_n(t) \over (t-t_0) },0] 
= -{ a_{-N}\over t_0^N}
-{ a_{-N}\over t_0^N}+\cdots
-{ a_{-1}\over t_0^1}
\equn
$$
Provided explicit formulae exist for the sub-leading singularities we
may in principle use Cauchy to compute the result.   The soft theorems
thus might be useful in constructing these amplitudes. 

Taking leg $n$ to be of positive helicity and scaled according to \eqref{appysoft},
for large $t$ we find 
$$
A: M_n(1^-,2^+,3^+,\cdots  , n^+)(t) \longrightarrow t^{n-4} 
\equn
$$
While taking leg $1$ to have negative helicity and scaled according to the complex conjugate of \eqref{appysoft},
at large $t$ we find
$$
B: M_n(1^-,2^+,3^+,\cdots  , n^+)(t) \longrightarrow t^{-3} 
\equn
$$

There is an unfortunate conspiracy here.  In case A, the shifted amplitude
has single poles except at $t=0$ AND at $t=0$ the soft-limit is well
known. However the large-$t$ behaviour generates a contribution at infinity which can not be directly obtained from a factorisation
of the amplitude.   In case B the large
$t$ limit is good however the amplitude does not satisfy the soft limit
theorems AND the amplitude has double poles at some points $t_i$ which
would require extra information to determine.  

\section{Mathematica form of Amplitudes}
\label{MathematicaAppendix}
At \url{http://pyweb.swan.ac.uk/~dunbar/graviton.html}  explicit forms of the single-minus amplitudes are 
available in {\tt Mathematica} format. These are presented as polynomials in $\lambda_i$ and
$\bar\lambda_i$, specifically as $\tt lambda[x,i]$ and $\tt barlambda[x,i]$.    The expressions satisfy:

$\bullet$ symmetry under interchange of positive legs.

$\bullet$ independence of the choice of positive shift leg.  This is actually quite a strenuous test which ties together all the different terms in the amplitude.
If we view amplitudes as the area of some polyhedron~\cite{Arkani-Hamed:2013jha} then the BCFW shift corresponds to a triangulation of the area.  

$\bullet$ The amplitudes all satisfy the leading soft behaviour: this checks (or fixes) the overall normalisations.

$\bullet$ The amplitudes have the correct collinear limits.
The collinear limit occurs when legs $k_a$ and $k_b$ are collinear,
$k_a\cdot k_b \longrightarrow 0$.  Unlike Yang-Mills
amplitudes, gravity amplitudes are not
singular in the collinear limit, but acquire a ``phase-singularity''~\cite{Bern:1998sv,Dunbar:2012aj}. 
If $k_a \longrightarrow z K$ and $k_b \longrightarrow (1-z) K $, the singularity is
\be
M_n( \cdots , a^{h_a}, b^{h_b} ) \coli{a}{b}   \sum_{h'}
\SP_{-h'}^{h_ah_b}  M_{n-1} (\cdots , K^{h'} ) 
\ee
where the $h$'s denote the various helicities of the
gravitons.  
The non-zero
``splitting functions'' are
\begin{align}
   \SP_{-}^{++}&= -{ \spb{a}.{b} \over z(1-z)  \spa{a}.{b}  }, 
\;\;\; 
  \SP_{+}^{-+} 
= -{ z^3\spb{a}.{b} \over (1-z) \spa{a}.{b} }.
\end{align}
An important result of ref.~\cite{Bern:1998sv} is that the splitting functions  do not obtain loop corrections.

We have made  the amplitudes available in  a very explicit, if rather cumbersome, polynomial form. The size of these polynomials grows rapidly with
the number of points. In Table~\ref{sixpointN4} we present the {\tt LeafCount} of these {\tt Mathematica} expressions as a rough indicator of the growth. 
The MHV tree and all-plus one-loop amplitudes are included for comparison. 
None of these has been optimised and they are all equally inefficient.

\begin{table}[h]
\renewcommand{\arraystretch}{1.5}
\centering\begin{tabular}{r|rrrrr}
\toprule 
  Amplitude  $\;\;$ & $\; n=4$        & n=5   & 
$n=6 $ &  $n=7$ & $n=8$ 
\\
\hline 
 $ M_n^{\rm tree} (-,-,+,\cdots,+)$  &  $144$      & $512$
& $2,318$ & $13,244$ & $95,000$ \\
$ M_n^{\rm 1-loop}(+,+,+,\cdots,+)$  & $216$      &
 $4,085$ & $84,470$ & $553,145$ & $3,814,189$ \\
$M_n^{\rm 1-loop}(-,+,+,\cdots,+) $  & $252$     &
 $\;\; 4,959$ &  $\;\;150,474$ & $\;\;4,462,636$ &  $\;\;220,910,118$\\
$\Delta_n$  &  $-$    &
 $-$ & $42,750$ & $943,051$ & $47,265,553$\\
 \bottomrule
\end{tabular}
\caption{{\tt LeafCount} of some rational amplitudes expressed as polynomials in $\lambda$ and $\bar\lambda$.}
\label{sixpointN4}
\end{table}

\end{document}